\documentstyle[12pt]{article}

\begin{document}

\author{Yishi Duan\thanks{%
Institute of Theoretical Physics, Lanzhou University, Lanzhou 730000,
P.R.China} \thanks{%
E-mail: ysduan{\it @}lzu.edu.cn} and Ying Jiang$^*$\thanks{%
Corresponding Author, E-mail: itp3{\it @}lzu.edu.cn}}
\title{Can torsion play a role in angular momentum conservation law? }
\date{}
\maketitle

\begin{abstract}
In Einstein-Cartan theory, by the use of the general Noether theorem, the
general covariant angular-momentum conservation law is obtained with the
respect to the local Lorentz transformations. The corresponding conservative
Noether current is interpreted as the angular momentum tensor of the
gravity-matter system including the spin density. It is pointed out that,
assuming the tetrad transformation given by eq. (15), torsion tensor can
not play a role in the conservation law of angular
momentum. \\

\end{abstract}

\section{Introduction}

Conservation laws of energy-momentum and angular momentum have been of
fundamental interest in gravitational physics\cite{1}. Using the vierbein
representation of general relativity, Duan ( one of the present author ) 
{\it et al }obtained a general covariant conservation law of energy-momentum
which overcomes the difficulties of other expressions\cite{2}. This
conservation law gives the correct quadrupole radiation formula of energy
which is in good agreement with the analysis of the gravitational damping
for the pulsar PSR1916-13\cite{3}. Also, from the same point of view, Duan
and Feng\cite{4} proposed a covariant conservation law of angular momentum
in Riemann space--time which does not suffer from the flaws of the others.%
\cite{5,6,7}

On the other hand, though the Einstein theory of general relativity has
succeeded in many respects, there is an essential difficulty in this theory:
we could not get a successful renormalized quantum gravity theory\cite{8}.
In order to find renormalized theories, many physicists\cite{9,10} have
studied this problem in its more general aspects, i.e. extending Einstein's
theory to Einstein-Cartan theory, which includes torsion\cite{11}.

As is well known, torsion is a slight modification of the Einstein's theory
of relativity\cite{12}, but is a generalization that appears to be necessary
when one tries to conciliate general relativity with quantum theory. Like
opening a Pandora's box, many works have been done in this region\cite{14,15}%
. Today, general relativity with non-zero torsion is a major contender for a
realistive generalization of the theory of gravitation.

About two decades ago, Hehl\cite{11} gave, in Einstein-Cartan theory, an
expression of the angular momentum conservation law which was worked out
from Noether's theorem, but in that expression, all quantities carried
Riemannian indices and the total angular momentum depended on the
coordinative choice which is not an observable quantity. Some physicists\cite
{9,11} investigated the same problem from the local Poincar\'e
transformation and presented another expression of conservation law which is
not general covariant, and they did not provide the superpotentials which is
much more important in conservation law, hence this theory cannot be said to
be a very satisfactory theory. Recently, Hammond\cite{17} obtained his
expression of the angular momentum conservation law, unfortunately this
theory also met with the difficulties mentioned above.

Several years ago, the general covariant energy-momentum conservation law in
general space-time has been discussed successfully by Duan et.al\cite{16}.
In this paper, we will study the angular momentum conservation law in
Einstein-Cartan theory via the vierbein representation. General relativity
without vierbein is like a boat without a jib--without these vital
ingredients the going is slow and progress inhibited. Consequently, vierbein
have grown to be an indispensable tool in many aspects of general
relativity. More important, it is relevant to the physical observability\cite
{20}. Based on the Einstein's observable time and space interval, we take
the local point of view that any measurement in physics is performed in the
local flat reference system whose existence is guaranteed by the equivalence
principle, i.e. an observable object must carries the indices of the
internal space. Thus, we draw the support from vierbein not only for
mathematical reasons, but also because of physical measurement consideration.

This paper is organized as follows: In section 2, we discuss the general
conservation laws in general case. In section 3, by making use of the
general Noether theorem, we obtain, in Einstein--Cartan theory, the general
covariant angular-momentum conservation law with the respect to the local
Lorentz transformations. The corresponding conservative Noether current is
interpreted as the angular momentum tensor of the gravity-matter system
including the spin density. In section 4, we give a brief summary of the
above discussion and point out that torsion tensor can not play a role in
the conservation law of angular momentum.

\section{Conservation law in general case}

The conservation law is one of the important essential problems in
gravitational theory. It is due to the invariance of Lagrangian
corresponding to some transformations. In order to study the covariant
angular momentum conservation law, it is necessary to discuss conservation
law by the Noether theorem in general case.

The action of a system is 
\begin{equation}
\label{general action}I=\int_M{\cal L}(\phi ^A,\phi _{\;,\mu }^A)\,d^4x, 
\end{equation}
where $\phi ^A$ are independent variables with general index $A$, $\phi
_{\;,\mu }^A=\partial _\mu \phi ^A$. If the action is invariant under the
infinitesimal transformation 
\begin{equation}
\label{x transformation}x^{^{\prime }\mu }=x^\mu +\delta x^\mu , 
\end{equation}
\begin{equation}
\label{phi transformation}\phi ^{^{\prime }A}(x^{^{\prime }})=\phi
^A(x)+\delta \phi ^A(x), 
\end{equation}
and $\delta \phi ^A$ is zero on the boundary of the four-dimensional volume $%
M$, then we can prove that there is the relation 
\begin{equation}
\label{noether theorem}\frac \partial {\partial x^\mu }({\cal L}\delta x^\mu
+\frac{\partial {\cal L}}{\partial \phi _{\;,\mu }^A}\delta _0\phi ^A)+[%
{\cal L}]_{\phi ^A}\delta _0\phi ^A=0, 
\end{equation}
where $[{\cal L}]_{\phi ^A}$ is%
$$
[{\cal L}]_{\phi ^A}=\frac{\partial {\cal L}}{\partial \phi ^A}-\partial
_\mu (\frac{\partial {\cal L}}{\partial \phi _{\;,\mu }^A}), 
$$
and $\delta _0\phi ^A$ is the Lie derivative of $\phi ^A$%
\begin{equation}
\label{delta0phi}\delta _0\phi ^A=\delta \phi ^A(x)-\phi _{\;,\mu }^A\delta
x^\mu . 
\end{equation}

If ${\cal L}$ is the total Lagrangian density of the system, there is $[%
{\cal L}]_{\phi ^A}=0$, the field equation of $\phi ^A$ with respect to $%
\delta I=0.$ From (\ref{noether theorem}) we know that there is a
conservation equation corresponding to transformations (\ref
{x
transformation}) and (\ref{phi transformation}) 
\begin{equation}
\label{noether conservation}\frac \partial {\partial x^\mu }({\cal L}\delta
x^\mu +\frac{\partial {\cal L}}{\partial \phi _{\;,\mu }^A}\delta _0\phi
^A)=0. 
\end{equation}
This is just the conservation law in general case. It must be pointed out
that if ${\cal L}$ is not the total Lagrangian density of the system, then
as long as the action of ${\cal L}$ remains invariant under transformations (%
\ref{x transformation}) and (\ref{phi transformation}), (\ref
{noether
theorem}) is still tenable. But (\ref{noether conservation}) is
not admissible now due to $[{\cal L}]_{\phi ^A}\neq 0.$

In gravitational theory with the vierbein as element fields we can separate $%
\phi ^A$ as $\phi ^A=(e_\mu ^a,\psi ^B),$ where $e_\mu ^a$ is the vierbein
field and $\psi ^B$ is an arbitrary tensor under general coordinate
transformation. When $\psi ^B$ is $\psi ^{\mu _1\mu _2...\mu _k}$, we can
always scalarize it by%
$$
\psi ^{a_1a_2...a_k}=e_{\mu _1}^{a_1}e_{\mu _2}^{a_2}...e_{\mu _k}^{a_k}\psi
^{\mu _1\mu _2...\mu _k}, 
$$
so we can take $\psi ^B$ as a scalar field under general coordinate
transformations. In later discussion we can simplify the equations by such a
choice.

\section{Angular momentum conservation law in Einstein--Cartan theory}

In Einstein-Cartan theory, the total action of the gravity$-$matter system
is expressed as\cite{11} 
\begin{equation}
\label{action}I=\int_M{\cal L}d^4x=\int_M({\cal L}_g{\cal +L}_m)d^4x, 
\end{equation}
$$
{\cal L}_g=\frac{c^4}{16\pi G}\sqrt{-g}R. 
$$
${\cal L}_g$ is the gravitational Lagrangian density, $R$ is the scalar
curvature of the Riemann--Cartan space--time. The matter part Lagrangian
density ${\cal L}_m$ take the form ${\cal L}_m={\cal L}_m(\phi ^A,D_\mu \phi
^A)$, where the matter field $\phi ^A$ belongs to some representation of
Lorentz group whose generators are $I_{ab}$ $(a,b=0,1,2,3)$ and $%
I_{ab}=-I_{ba}$, $D_\mu $ is the covariant derivative operator of $\phi ^A$%
$$
D_\mu \phi ^A=\partial _\mu \phi ^A-\frac 12\omega _{\mu
ab}(I_{ab})_{\;B}^A\phi ^B. 
$$

As in E$-$C theory, the affine connection $\Gamma _{\mu \nu }^\lambda $ is
not symmetry in $\mu $ and $\nu $, i.e. there exists non-zero torsion tensor%
$$
T_{\;\mu \nu }^\lambda =\Gamma _{\mu \nu }^\lambda -\Gamma _{\nu \mu
}^\lambda . 
$$
It is well known, for vierbein field $e_\mu ^a$, the total covariant
derivative is equal to zero, i.e. 
\begin{equation}
\label{total derivative}{\cal D}_\mu \,e_\nu ^a\equiv \partial _\mu \,e_\nu
^a-\omega _{\mu ab}\,e_\nu ^b-\Gamma _{\mu \nu }^\lambda \,e_\lambda ^b=0.
\end{equation}
This formula is also can be looked upon as the definition of the spin
connection $\omega _{\mu ab}$. From (\ref{total derivative}), we can get the
total expansion of the spin connection with vierbein and torsion 
$$
\omega _{abc}=\bar \omega _{abc}+\frac 12(T_{cab}+T_{acb}+T_{bca}), 
$$
\begin{equation}
\bar \omega _{abc}=e_a^\mu (\partial _\mu e_b^\nu +\{_{\mu \sigma }^\nu
\}e_b^\sigma )e_{\nu c},
\end{equation}
where $\omega _{abc}=e_a^\mu \omega _{\mu bc},\;T_{abc}=e_{\lambda
a}\,e_b^\mu \,e_c^\nu T_{\;\mu \nu }^\lambda $ are, respectively, the
representation of spin connection and torsion tensor in vierbein theory, $%
e_a^\mu $ is the inverse of $e_\mu ^a$, $\{_{\mu \sigma }^\nu \}$ is the
Christoffel symbol. By tedious calculation, such decomposition of $\omega
_{abc}$ allows us to obtain the identity\cite{16} 
\begin{equation}
\label{lagrangian 1}{\cal L}_g=\frac{c^4}{16\pi G}\sqrt{-g}R=\frac{c^4}{%
16\pi G}\Delta -{\cal L}_{\bar \omega }+{\cal L}_T-\frac{c^4}{8\pi G}{\cal L}%
_{\partial T},
\end{equation}
where 
\begin{equation}
\label{lagrangian 2}{\cal L}_{\bar \omega }=\frac{c^4}{16\pi G}\sqrt{-g}%
(\bar \omega _a\bar \omega _a-\bar \omega _{abc}\bar \omega
_{cba}),\;\;\;\;\bar \omega _a=\bar \omega _{bab},
\end{equation}
\begin{equation}
\label{lagrangian 3}\Delta =\partial _\mu (\sqrt{-g}(e_a^\mu \partial _\nu
e_a^\nu -e_a^\nu \partial _\nu e_a^\mu ),
\end{equation}
\begin{equation}
\label{lagrangian 4}{\cal L}_T=\frac{c^4}{16\pi G}\sqrt{-g}(T_aT_a-\frac
12T_{abc}T_{cba}-\frac 14T_{abc}T_{abc}),\;\;T_a=T_{bab},
\end{equation}
\begin{equation}
\label{lagrangian 5}{\cal L}_{\partial T}=\partial _\mu (\sqrt{-g}e_a^\mu
T_a).
\end{equation}

It is well known that in deriving the general covariant conservation law of
energy momentum in general relativity, the general displacement
transformation, which is a generalization of the displacement transformation
in the Minkowski space-time, was used\cite{18}. In the local Lorentz
reference frame, the general displacement transformation takes the same form
as that in the Minkowski space-time. This implies that general covariant
conservation laws are corresponding to the invariance of the action under
local transformations. We may conjecture that since the conservation law for
angular momentum in special relativity corresponds to the invariance of the
action under the Lorentz transformation, the general covariant conservation
law of angular momentum in general relativity may be obtained by means of
the local Lorentz invariance.

We choose vierbein $e_\mu ^a$, torsion $T_{abc}$ and the matter field $\phi
^A$ as independent variables. Under the local Lorentz transformation 
\begin{equation}
\label{transformation 1}e_\mu ^a(x)\rightarrow e_\mu ^{\prime a}(x)=\Lambda
_{\;\,b}^a(x)e_\mu ^b(x),\;\;\;\Lambda _{\;\,c}^a(x)\Lambda
_{\;\,b}^c(x)=\delta _b^a, 
\end{equation}
$T_{abc}$ and $\phi ^A$ transform as 
\begin{equation}
\label{transformation 2}T_{abc}\rightarrow T_{abc}^{\prime }(x)=\Lambda
_{\;\,c}^l(x)\Lambda _{\;\,c}^m(x)\Lambda _{\;\,c}^n(x)T_{lmn}, 
\end{equation}
\begin{equation}
\label{transformation 3}\phi ^A\rightarrow \phi ^{\prime A}(x)=D(\Lambda
(x))_{\;\,B}^A\phi ^B(x). 
\end{equation}
Since the coordinates $x^\mu $ do not transform under the local Lorentz
transformation, $\delta x^\mu =0$, from (\ref{delta0phi}), it can be proved
that in this case, $\delta _0\rightarrow \delta $. It is required that $%
{\cal L}_m$ is invariant under (\ref{transformation 1}) and ${\cal L}_g$ is
invariant obviously. So under the local Lorentz transformation (\ref
{transformation 1}) ${\cal L}$ is invariant. In the light of the discussion
in section 2, we would like to have the relation 
\begin{equation}
\label{1}\frac \partial {\partial x^\mu }(\frac{\partial {\cal L}}{\partial
\partial _\mu e_a^\nu }\delta e_a^\nu +\frac{\partial {\cal L}}{\partial
\partial _\mu \phi ^A}\delta \phi ^A+\frac{\partial {\cal L}}{\partial
\partial _\mu T_{abc}}\delta T_{abc})+[{\cal L}]_{e_a^\nu }\delta e_a^\nu +[%
{\cal L}]_{T_{abc}}\delta T_{abc}+[{\cal L}]_{\phi ^A}\delta \phi ^A=0, 
\end{equation}
where $[{\cal L}]_{e_a^\nu }$, $[{\cal L}]_{T_{abc}}$ and $[{\cal L}]_{\phi
^A}$ are the Euler expressions defined as%
$$
[{\cal L}]_{e_a^\nu }=\frac{\partial {\cal L}}{\partial e_a^\nu }-\partial
_\mu \frac{\partial {\cal L}}{\partial \partial _\mu e_a^\nu }, 
$$
$$
[{\cal L}]_{T_{abc}}=\frac{\partial {\cal L}}{\partial T_{abc}}-\partial
_\mu \frac{\partial {\cal L}}{\partial \partial _\mu T_{abc}}, 
$$
$$
[{\cal L}]_{\phi ^A}=\frac{\partial {\cal L}}{\partial \phi ^A}-\partial
_\mu \frac{\partial {\cal L}}{\partial \partial _\mu \phi ^A}. 
$$
Using the Einstein equation $[{\cal L}]_{e_a^\nu }=0$, Einstein-Cartan
equation $[{\cal L}]_{T_{abc}}=0$ and the equation of motion of matter $[%
{\cal L}]_{\phi ^A}=0$, we get following by (\ref{1}) 
\begin{equation}
\label{2}\frac \partial {\partial x^\mu }(\frac{\partial {\cal L}_g}{%
\partial \partial _\mu e_a^\nu }\delta e_a^\nu +\frac{\partial {\cal L}_g}{%
\partial \partial _\mu T_{abc}}\delta T_{abc})+\frac \partial {\partial
x^\mu }(\frac{\partial {\cal L}_m}{\partial \partial _\mu e_a^\nu }\delta
e_a^\nu +\frac{\partial {\cal L}_m}{\partial \partial _\mu \phi ^A}\delta
\phi ^A)=0, 
\end{equation}
where we have used the fact that only ${\cal L}_m$ contains the matter field 
$\phi ^A,$ and with its structure ${\cal L}_m$ does not possess $\partial
_\mu T_{abc}$. Consider the infinitesimal local Lorentz transformation $%
\Lambda _{\;b}^a(x)=\delta _{\;b}^a+\alpha _{\;b}^a(x),\;\alpha
_{ab}=-\alpha _{ba}$, $D(\Lambda )$ can be linearized as $[D(\Lambda
)]_{\;B}^A=\delta _{\;B}^A+\frac 12(I_{ab})_{\;B}^A\alpha _{ab}$, we have%
$$
\delta e_a^\nu (x)=\alpha _{ab}e_b^\nu (x), 
$$
\begin{equation}
\label{4}\delta T_{abc}(x)=\alpha _{ad}T_{dbc}(x)+\alpha
_{bd}T_{adc}(x)+\alpha _{cd}T_{abd}(x), 
\end{equation}
$$
\delta \phi ^A(x)=\frac 12(I_{ab})_{\;B}^A\phi ^B(x)\alpha _{ab}(x). 
$$

We introduce $j_{ab}^\mu $%
\begin{equation}
\sqrt{-g}j_{ab}^\mu \alpha _{ab}=\frac 3c[\frac{\partial {\cal L}_{\bar
\omega }}{\partial \partial _\mu e_a^\nu }e_b^\nu \alpha _{ab}-\frac{%
\partial {\cal L}_m}{\partial \partial _\mu e_a^\nu }e_b^\nu \alpha
_{ab}-\frac 12\frac{\partial {\cal L}_m}{\partial \partial _\mu \phi ^A}%
(I_{ab})_{\;B}^A\phi ^B\alpha _{ab}], 
\end{equation}
then ($\ref{2}$) can be rewritten as 
\begin{equation}
\label{3}\partial _\mu (\sqrt{-g}j_{ab}^\mu \alpha _{ab})-\frac{3c^3}{16\pi G%
}\partial _\mu (\frac{\partial \Delta }{\partial \partial _\mu e_a^\nu }%
e_b^\nu \alpha _{ab})+\frac{3c^3}{8\pi G}\partial _\mu (\frac{\partial {\cal %
L}_{\partial T}}{\partial \partial _\mu e_a^\nu }e_b^\nu \alpha _{ab}+\frac{%
\partial {\cal L}_{\partial T}}{\partial \partial _\mu T_{abc}}\delta
T_{abc})=0. 
\end{equation}
From (\ref{lagrangian 3}) one can get easily that 
\begin{equation}
\label{delta}\frac{\partial \Delta }{\partial \partial _\mu e_a^\nu }e_b^\nu
\alpha _{ab}=\alpha _{ab}\partial _\mu (\sqrt{-g}V_{ab}^{\mu \lambda }), 
\end{equation}
where 
\begin{equation}
\label{super potential}V_{ab}^{\mu \lambda }=e_a^\mu e_b^\lambda -e_b^\mu
e_a^\lambda . 
\end{equation}
Now, let us investigate the third term in left-hand side (LHS) of ($\ref{3}$%
). With ($\ref{lagrangian 5}$), we can get 
\begin{equation}
\label{5}\frac{\partial {\cal L}_{\partial T}}{\partial \partial _\mu
e_a^\nu }e_b^\nu \alpha _{ab}=\sqrt{-g}T_ae_b^\mu \alpha _{ab}, 
\end{equation}
and from ($\ref{lagrangian 5}$) and ($\ref{4}$), we also have 
\begin{equation}
\label{6}\frac{\partial {\cal L}_{\partial T}}{\partial \partial _\mu T_{abc}%
}\delta T_{abc}=\sqrt{-g}e_a^\mu T_b\alpha _{ab}. 
\end{equation}
Considering that $\alpha _{ab}$ is antisymmetry, i.e. $\alpha _{ab}=-\alpha
_{ba}$, we draw the conclusion from (\ref{5}) and (\ref{6}) that the third
term in LHS of (\ref{3}) is equal to zero. Substituting (\ref{delta}), (\ref
{5}) and (\ref{6}) into (\ref{3}), we obtain 
\begin{equation}
\partial _\mu (\sqrt{-g}j_{ab}^\mu )\alpha _{ab}+[\sqrt{-g}j_{ab}^\mu -(%
\frac{3c^3}{16\pi G})\partial _\nu (\sqrt{-g}V_{ab}^{\nu \mu })]\partial
_\mu \alpha _{ab}=0. 
\end{equation}
Since $\alpha _{ab}$ and $\partial _\mu \alpha _{ab}$ are independent of
each other, we must have 
\begin{equation}
\label{j}\partial _\mu (\sqrt{-g}j_{ab}^\mu )=0, 
\end{equation}
\begin{equation}
\label{pj}j_{ab}^\mu =\frac{3c^3}{16\pi G}\frac 1{\sqrt{-g}}\partial _\nu (%
\sqrt{-g}V_{ab}^{\nu \mu }), 
\end{equation}
or 
\begin{equation}
\label{expand}j_{ab}^\mu =\frac{3c^3}{16\pi G}(\bar \omega _ae_b^\mu +\bar
\omega _{abc}e_c^\mu -\bar \omega _be_a^\mu -\bar \omega _{bac}e_c^\mu ). 
\end{equation}
From (\ref{j}) and (\ref{pj}), it can be concluded that $j_{ab}^\mu $ is
conserved identically. As usual, we call $V_{ab}^{\nu \mu }$
superpotentials. Since the current $j_{ab}^\mu $ is derived from the local
Lorentz invariance of the total Lagrangian, it can be interpreted as the
total angular momentum tensor density of the gravity-matter system, and it
contains the spin density of the matter field: $(\partial {\cal L}%
_m)/(\partial \partial _\mu e_a^\nu ).$ From the above discussion, we see
that not only the current $j_{ab}^\mu $ but also the superpotential $%
V_{ab}^{\mu \nu }$ do not have any terms relevant to torsion tensor, all of
them are only determined by the vierbein.

For a globally hyperbolic Riemann-Cartan manifold, there exist Cauchy
surfaces $\Sigma _t$ foliating $M$. We choose a submanifold $D$ of $M$
joining any two Cauchy surfaces $\Sigma _{t_1}$ and $\Sigma _{t_2}$ so the
boundary $\partial D$ of $D$ consists of three parts $\Sigma _{t_1},\Sigma
_{t_2}$ and $A$ which is at spatial infinity. For an isolated system, the
space-time should be asymptotically flat at spatial infinity, so the
vierbein have the following asymtotical behavior\cite{19} 
\begin{equation}
\label{ee}\lim _{r\rightarrow \infty }(\partial _\mu e_{\nu a}-\partial _\nu
e_{\mu a})=0 
\end{equation}
Since 
$$
\sqrt{-g}V_{ab}^{\mu \nu }=\frac 12\epsilon ^{\mu \nu \lambda \rho }\epsilon
_{abcd}e_{\lambda c}e_{\rho d}, 
$$
we have $\lim _{r\rightarrow \infty }\partial _\lambda (\sqrt{-g}%
V_{ab}^{\lambda \mu })=0$. Thus, we can get the total conservative angular
momentum from (\ref{j}) and (\ref{pj}) 
\begin{equation}
J_{ab}=\int_{\Sigma _t}j_{ab}^\mu \sqrt{-g}d\Sigma _\mu =\frac{3c^3}{16\pi G}%
\int_{\partial \Sigma _t}\sqrt{-g}V_{ab}^{\mu \nu }d\sigma _{\mu \nu }, 
\end{equation}
where $\sqrt{-g}d\Sigma _\mu $ is the covariant surface element of $\Sigma
_t $, $d\Sigma _\mu =\frac 1{3!}\epsilon _{\mu \nu \lambda \rho }dx^\nu
\wedge dx^\lambda \wedge dx^\rho $, $d\sigma _{\mu \nu }=\frac 12\epsilon
_{\mu \nu \lambda \rho }dx^\lambda \wedge dx^\rho .$

\section{Discussion}

In summary, we have succeeded in obtaining an expression of an angular
momentum conservation law in Riemann--Cartan space--time. This conservation
law has the following main properties:

1. It is a covariant theory with respect to the generalized coordinate
transformations, but the angular momentum tensor is not covariant under the
local Lorentz transformation which, due to the equivalent principle, is
reasonable to require.

2. For a closed system, the total angular momentum does not depend on the
choice of the Riemannian coordinates and, according (\ref{ee}), the
space-time at spatial infinity is flat, thus the conservative angular
momentum $J_{ab}$ should be a covariant object when we make a Lorentz
transformation $\Lambda _{ab}=A_{ab}=$ constant at spatial infinity, as in
special relativity%
$$
J_{ab}^{^{\prime }}=A_{\;a}^cA_{\;b}^dJ_{cd}. 
$$
To understand this the key point is that to obtain $J_{ab}$, one has to
enclose everything of the closed system, and every point of space-time at
spatial infinity belongs to the same Minkowski space-time in that region.
This means that in general relativity for a closed system, the total angular
momentum $J_{ab}$ must be looked upon as a Lorentz tensor like that in
spatial relativity.

3. The conservative angular-momentum current and the corresponding
superpotential in Einstein--Cartan theory are the same with those in
Einstein theory\cite{4}, torsion can not play a role in the conservation
law. Both angular-momentum current and the superpotential are determined
only by vierbein field. This result makes us be flabbergasted. As is well
known, the torsion tensor is closely related to the spin density tensor
through the famous Einstein--Cartan field equation, it is natural for us to
believe that the torsion tensor should play some important role in the
conservation law. Is that true? As mentioned in this paper, with the
accurate calculation, the answer is no. Then a question arises: Why can not
the torsion play a role in the angular momentum conservation law? We hope we
can work it out in the future.

\end{document}